\documentclass[showpacs, pra,twocolumn,preprintnumbers ,amsmath, amssymb, superscriptaddress, aps]{revtex4-2}
\usepackage{amsfonts}
\usepackage{color}
\usepackage{amsmath,amssymb}
\usepackage{pifont}
\usepackage{amssymb}  
\usepackage{bbold}
\usepackage{float}
\usepackage{subfloat}

\usepackage[caption=false]{subfig}
\usepackage{tikz}
\usepackage{makecell}
\usepackage{subfig}
\usepackage{pifont}   
\usepackage{graphicx} 
\graphicspath{{Figures/}}
\usepackage{dcolumn}  
\usepackage{bm}       
\usepackage{multirow} 
\usepackage{placeins}
\usepackage[colorlinks]{hyperref}
\usepackage{mathtools}
\usepackage{appendix}

\def \be{\begin{align}}
	\def \ee{\end{align}}
\def \bea{\begin{eqnarray}}
	\def \eea{\end{eqnarray}}

\begin{document}

        \title{
        {Transport properties in ABC-ABA-ABC trilayer graphene junctions}}
        \date{\today}
          \author{Abderrahim El Mouhafid}
        \email{elmouhafid.a@ucd.ac.ma}
        \affiliation{Laboratory of Theoretical Physics, Faculty of Sciences, Choua\"ib Doukkali University, PO Box 20, 24000 El Jadida, Morocco}
         \author{Mouhamadou Hassane Saley}
        \email{hassanesaley.m@ucd.ac.ma}
        \affiliation{Laboratory of Theoretical Physics, Faculty of Sciences, Choua\"ib Doukkali University, PO Box 20, 24000 El Jadida, Morocco}
     \author{Ahmed Jellal}
        \email{a.jellal@ucd.ac.ma}
        \affiliation{Laboratory of Theoretical Physics, Faculty of Sciences, Choua\"ib Doukkali University, PO Box 20, 24000 El Jadida, Morocco}
        \affiliation{Canadian Quantum Research Center, 204-3002 32 Ave Vernon,  BC V1T 2L7, Canada}

        
\begin{abstract} 
	
	Trilayer graphene {(TLG)} consists of three layers of graphene arranged in a particular stacking order. In the case of ABC-ABA-ABC stacking, the layers are arranged in an A-B-C sequence, followed by an A-B-A sequence, and again an A-B-C sequence. This stacking arrangement introduces specific electronic properties and band structures due to the different stacking configurations.
	We focus on elucidating the transport properties of a p-n-p junction formed with ABC-ABA-ABC stacking TLG. Employing the transfer matrix method and considering continuity conditions at the junction boundaries, we establish transmission and reflection probabilities, along with conductance. Notably, electron transport through the ABC-ABA-ABC junction exhibits Klein tunneling, resulting in substantial conductance even in the absence of a potential barrier $V_0$. This effect arises from the effective barrier induced by our specific stacking, facilitating the passage of a maximal number of electrons. However, the presence of $V_0$ diminishes Klein tunneling, leading to conductance minima. Furthermore, our findings highlight that interlayer bias $\delta$ induces a hybridization of the linear and parabolic bands of ABA-TLG within the junction, reducing resonances. In cases where $\delta\neq0$ and $V_0\neq0$, we observe a suppression of the gap, contrary to the results obtained in ABC tunneling studies where a gap exists.


\end{abstract}          
        
\maketitle

\section{Introduction}

Graphene, a gapless semiconductor composed of a single layer of carbon atoms, exhibits massless charge carriers with a linear energy band, governed by the Dirac equation for relativistic particles \cite{graphene, Novoselov, Zhang, Morozov, Geim}. This unique characteristic has sparked considerable interest among researchers, theorists, and experimenters. For instance, in graphene, the electron transport across a potential barrier is marked by perfect transmission, known as Klein tunneling \cite{Katsnelson, Young}. Additionally, graphene displays an unconventional quantum Hall effect \cite{Novoselov, Zhang, Gusynin}. Moreover, graphene possesses transparency \cite{Nair}, high electronic mobility \cite{Morozov, Shishir}, impressive thermal conductivity \cite{Balandin, Ghosh}, and a remarkable strength of 130 GPa \cite{Lee, Papageorgiou}. These diverse properties underscore the significance of graphene across a broad spectrum of applications. 


In addition to monolayer graphene (MLG), multilayer graphene, involving the stacking of two or more layers, exhibits appealing properties dependent on the stacking type and layer count. Specifically, Bernal (AB) bilayer graphene (BLG) features four parabolic energy bands, allowing for gap opening through the application of a perpendicular electric field \cite{Edward,McCann,Tang,Van}. Conversely, AA-BLG displays a linear energy dispersion akin to MLG but with a shift in the positions of the two Dirac cones \cite{Lobato,Matthew}. Unlike AB-BLG, the application of an electric field in AA-BLG does not result in a band gap \cite{Tabert,Ezzi}. Furthermore, AA-BLG exhibits Klein tunneling similar to MLG, while AB-BLG demonstrates perfect reflection \cite{Matthew,Ezzi,Bahlouli,Katsnelson,Tudorovskiy,GU,Peeters,Hassane}. Trilayer graphene comes in three distinct configurations: ABC (rhombohedral), ABA (bernal), and AAA stackings \cite{Guinea,Aoki,Craciun,Jing,Lui,Campos,Sena,Cheng,Sugawara,ELMouhafid2200308,Moradian,Bahaoui}. The ABC stacking displays a cubic dispersion with conduction and valence bands touching at low energy, while the ABA stacking exhibits both MLG linear bands and BLG parabolic bands \cite{Guinea,Jing,Campos,Khodasa}. The AAA stacking comprises three copies of MLG linear bands \cite{Bahaoui}. Studies have shown that the presence of an electric field breaks the inversion symmetry of ABC-TLG, leading to the appearance of a band gap, while ABA-TLG remains metallic \cite{Lui,Cheng,Fan,LeRoy,Zou}. In ABA-TLG, the mirror symmetry is broken in the presence of an electric field, resulting in the hybridization of the linear and parabolic bands \cite{Koshino}.

In a recent experimental investigation, Yin {\it et al.} documented the emergence of a domain wall in the transitional zone between ABA and ABC stacking \cite{Yin}. Building on this research, we explore electron tunneling through an ABC-ABA-ABC junctions with the goal of evaluating its impact on the transport properties of Dirac fermions. Our analysis involves determining transmission and reflection probabilities along with the corresponding conductance. In scenarios without a potential barrier ($V_0$), we demonstrate that electrons can undergo perfect tunneling through the ABC-ABA-ABC junction under normal incidence, a phenomenon known as Klein tunneling. As a result, maximal conductance is achieved, attributed to the junction inducing an effective potential that facilitates the optimal flow of electrons, leading to total transmission. However, the introduction of a potential barrier diminishes Klein tunneling, as evident in the observed minima in conductance. Furthermore, we show that the presence of an interlayer bias ($\delta$) significantly reduces transmission in specific regions, primarily due to the hybridization of MLG-like and AB-BLG-like bands in ABA-TLG. Notably, contrary to the findings in \cite{vanduppen195439}, no gap is observed for $V_0\neq0$ and $\delta \neq 0$. Our study includes various comparisons of these results with findings related to the p-n-p junction (potential barrier) using either ABC-TLG or ABA-TLG exclusively.

The present paper is organized as follows. In Sec. \ref{Model}, we introduce the theoretical model for our system and calculate the transmission and reflection probabilities. The numerical results obtained are presented and analyzed in Sec. \ref{RRDD}. Ultimately, a summary of the key findings is presented in Sec. \ref{CC}.


\section{Model and method}\label{Model}

We recall that trilayer graphene is characterized by a specific arrangement of carbon atoms forming a structure composed of three layers of single-layer or monolayer graphene sheets. Each monolayer exhibits a hexagonal crystal structure with atoms labeled A and B in its unit cell, and their interatomic distance is $a_0=0.142$ nm \cite{Partoens2007}, with an intra-layer coupling of approximately $\gamma_0\approx3$ eV \cite{Zhang2011}.
The trilayer structure consists of a first monolayer followed by a second monolayer, aligning with the first in an alternating pattern known as ABA stacking. In ABA stacking, the second monolayer is directly positioned on top of the first monolayer, and the third monolayer aligns with the first monolayer in the same pattern, creating a repeating ABA arrangement throughout the trilayer. Alternatively, trilayer graphene can adopt another stacking configuration called ABC stacking. In ABC stacking, the carbon atoms in the third monolayer are aligned with those in the second monolayer in a different pattern. Specifically, in ABC stacking, the third monolayer is rotated by $60^{\circ}$  relative to the first monolayer, introducing a stagger between the layers.

  
\subsection{Hamiltonian model}

\begin{figure}[ht]
 	\centering
 	\includegraphics[width=3 in]{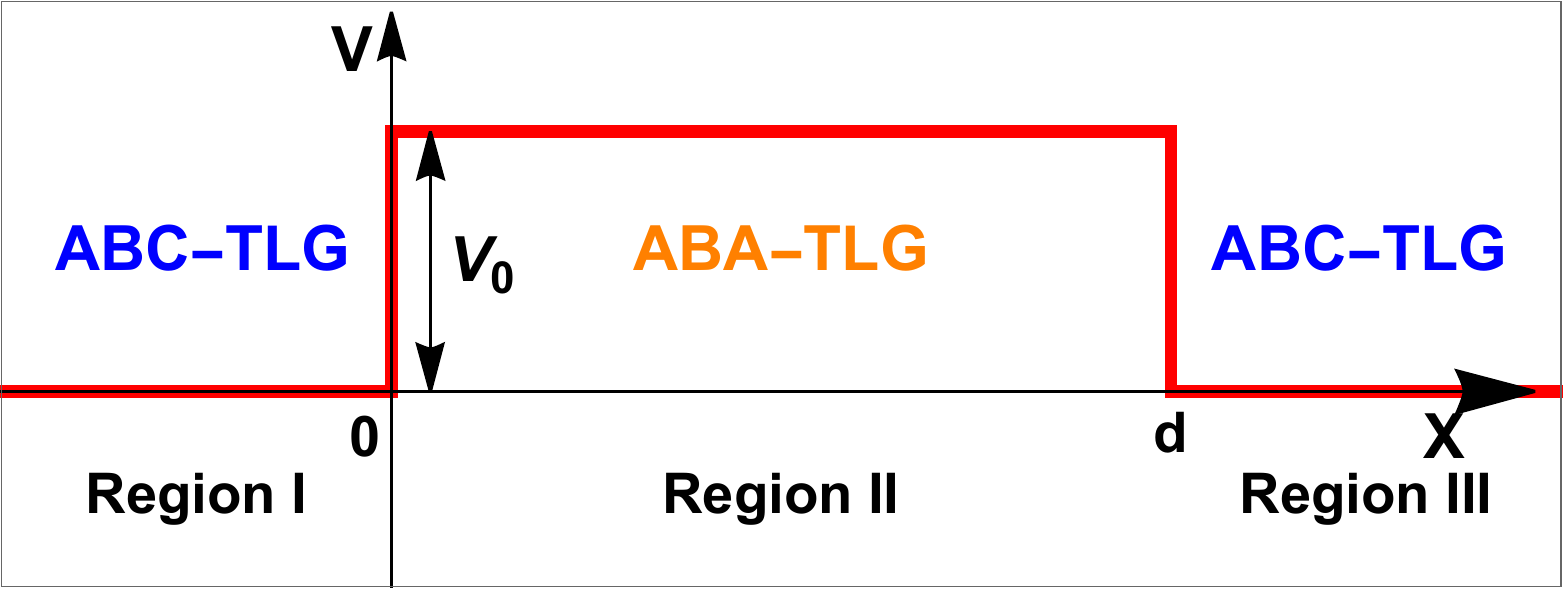} \caption{(Color online) Profile illustrating the trilayer graphene junction in an ABC-ABA-ABC configuration, along with a potential barrier characterized by a height $V_0$ and a width  $d$.}\label{Shematicbarrier}
 \end{figure}

Fig. \ref{Shematicbarrier} depicts two regions of ABC-TLG ($j$=I, III) characterized by zero potential, alongside an intermediate region of ABA-TLG ($j$=II) with a nonzero potential. The potential on the top and bottom layers is defined as
 \begin{align}
 	V_{j}(x)=\left\{\begin{array}{l}
 		V_0\pm \delta, \ \ \ j=\text{II}\\
 	0, \ \ \ \ \ \ \ \ \ \ j=\text{I, III}
 	\end{array}
 		\right.
 \end{align} 
 where $V_0$ represents the height of the potential barrier  and $\delta$ denotes the electrostatic potential difference (bias). 
 To obtain the solution within each region $j$, we solve the eigenvalue equation $H_j\psi_j=E_j\psi_j$, where the effective Hamiltonians $H_j$ near the Dirac point $K$ for ABC-TLG (regions I, III) and ABA-TLG (region II), respectively, can be derived from the continium limit of the tight-binding model \cite{Fan,vanduppen195439, Partoens2007}
\begin{align}
& \mathcal{H}_{ABC}(\vec p)=
\begin{pmatrix}
\mathcal{H}_1(\vec p) & \Gamma & 0 \\
\Gamma^{\dag } & \mathcal{H}_2(\vec p) & \Gamma \\
0 & \Gamma^{\dag} & \mathcal{H}_3(\vec p)%
\end{pmatrix}\label{HamABC},
\\
&\mathcal{H}_{ABA}(\vec p)=
\begin{pmatrix}
\mathcal{H}_1(\vec p) & \Gamma & 0 \\
\Gamma^{\dag } & \mathcal{H}_2(\vec p) & \Gamma^{\dag } \\
0 & \Gamma & \mathcal{H}_3(\vec p)%
\end{pmatrix}%
\label{HamABA1},
\end{align}
where the interlayer coupling $\Gamma$ is  given by
\begin{align}
\Gamma=
\begin{pmatrix}
0 & 0 \\
\gamma_{1} & 0%
\end{pmatrix}%
\label{Gamma2},
\end{align}
and $\mathcal{H}_i(\vec p)=\hbar v_F \vec\sigma\cdot\vec p+V_{i}(x)\mathbb{I}_2$ is the SLG Hamiltonian of the \textit{i}-th layer $(i=1,2,3)$, with $\mathbb{I}_2$ is a unit matrix, $v_F\approx1.01\times10^6$m/s is the Fermi velocity, $\vec\sigma=(\sigma_x,\sigma_y)$ are the Pauli matrices, $\vec p=(p_{x},p_{y})$ denotes  the in-plane momentum, 
$\gamma_{1}=0.4$ eV is the nearest neighbor coupling term between adjacent layers. 
These Hamiltonians are formulated based on the eigenfunctions of atomic orbitals
\begin{equation}
\Psi_{ABC/ABA} =\left( \psi _{A_{1}},\psi _{B_{1}},\psi _{A_{2}},\psi _{B_{2}},\psi_{A_{3}},\psi _{B_{3}}\right)^{\dagger},  \label{}
\end{equation}
where the indices signify the sublattice linked to the corresponding eigenfunction and the symbol $\dagger$ stands for transpose. 

{We should clarify that we are studying a simplified or toy model, which is characterized by a simplification that makes it more tractable. That is, while it includes many important factors such as interlayer couplings and dispersions near the $K$ valley, it assumes that the barriers between different lattice regions lead to negligible intervally scattering. We note that realistically shifts between lattice structures occur on the scale of a lattice constant, and therefore momentum transfers could potentially connect different $K$ points - to keep the model more tractable. Our model, as a simplification, neglects contributions from this possible effect. A more detailed and extensive study would be necessary to fully characterize the importance of phenomena that could lead to tunneling between different $K$ points, and to fully understand when our simplified model is fully accurate despite the possibility of intervally scattering.}

A simplification can be achieved by converting the Hamiltonian of ABA-TLG (\ref{HamABA1}) into a block-diagonal structure through a unitary transformation that symmetrically and antisymmetrically combines the orbital eigenfunctions, as outlined in \cite{Koshino}. This transformation can be expressed as
\begin{equation}\label{}
U=\frac{1}{\sqrt{2}}
\begin{pmatrix}
  1 & 0 & 0 & 0 & -1 & 0 \\
  0 & 1 & 0 & 0 & 0 & -1\\
  1 & 0 & 0 & 0 & 1 & 0 \\
  0 & 1 & 0 & 0 & 0 & 1 \\
  0 & 0 & \sqrt{2} & 0 & 0 & 0 \\
  0 & 0 & 0 & \sqrt{2} & 0 & 0 \\
\end{pmatrix},
\end{equation}
which leads to the following Hamiltonian
\begin{align}
\mathcal{H}_{ABA}(\vec p)=\begin{pmatrix}
\mathcal{H}_1(\vec p) & \delta \mathbb{I}_2 & 0 \\
\delta \mathbb{I}_2 & \mathcal{H}_2(\vec p) & \sqrt{2}\Gamma \\
0 & \sqrt{2}\Gamma^{\dag } & \mathcal{H}_3(\vec p)%
\end{pmatrix}%
\label{HamABA2}.
\end{align}
We consider the domain wall oriented in the $y$-direction with infinite length. Consequently, the system exhibits translational invariance, and the momentum $p_y$ is conserved. This allows us to formulate the eigenfunctions for ABC ({$j=$I, III}) and ABA ($j=\text{II}$), respectively, as
\begin{align}
&\psi_j(x,y) =e^{ik_y y}({\phi}_{A_{1}j
},{\phi}_{B_{1}j},{\phi}_{A_{2}j},{\phi}_{B_{2}j},{\phi}_{A_{3}j},{\phi}_{B_{3}j})^{\dag}, \label{wavefABC}\\
&\psi_{\text{II}}(x,y) =e^{ik_y y}({\chi}_{A_{1}
},{\chi}_{B_{1}},{\chi}_{A_{2}},{\chi}_{B_{2}},{\chi}_{A_{3}},{\chi}_{B_{3}})^{\dag}. \label{wavefABA}
\end{align}
To streamline the notation, we introduce the length scale  $a=\frac{\hbar v_{F}}{\gamma_{1}}\approx 1.64$  nm, representing the interlayer coupling length, 
together with the dimensionless quantities: $E_j \rightarrow\ \frac{E_j}{\gamma_1}$, {$V_0 \rightarrow\ \frac{V_0}{\gamma_1}$}, $\delta_j \rightarrow\ \frac{\delta_j}{\gamma_1}$, {$k_y \rightarrow\ ak_y$} and {$\vec r \rightarrow\ \frac{\vec r}{a}$.}

\subsection{Solutions of energy spectrum}

\subsubsection{ABC-TLG in regions {\textnormal{I}} and  {\textnormal{III}}}

Concerning regions I and III, the eigenvalue equation associated with the Hamiltonian (\ref{HamABC}) leads to a system of six-coupled differential equations
\begin{subequations}\label{eqdeffABC}
 \begin{align}
&-i(\partial_{x}+k_{y})\phi_{B_{1}j} =E_j\phi_{A_{1}j},
 \label{eqsABCd1} \\
&-i(\partial_{x}-k_{y})\phi_{A_{1}j} =E_j\phi_{B_{1}j}-\phi_{A_{2}j}, \label{eqsABCd2}  \\
&-i(\partial_{x}+k_{y})\phi_{B_{2}j}=E_j\phi_{A_{2}j}-\phi_{B_{1}j}, \label{eqsABCd3}  \\
&-i(\partial_{x}-k_{y})\phi_{A_{2}j}=E_j\phi_{B_{2}j}-\phi_{A_{3}j},
\label{eqsABCd4} \\
&-i(\partial_{x}+k_{y})\phi_{B_{3}j}=E_j\phi_{A_{3}j}-\phi_{B_{2}j},
\label{eqsABCd5} \\
&-i(\partial_{x}-k_{y})\phi_{A_{3}j}=E_j\phi_{B_{3}j}
 \label{eqsABCd6},
\end{align}
\end{subequations}
with $k_{y}$ represents the wave vector in the $y$-direction. To uncouple these equations, let us initially derive $\phi_{A_1j}$ and $\phi_{B_3j}$ in relation to $\phi_{B_1j}$ and $\phi_{A_3j}$ using Eqs. \eqref{eqsABCd1} and \eqref{eqsABCd6}. Subsequently, we substitute these expressions into the second and fifth equations, respectively, yielding
\begin{subequations}\label{}
 \begin{eqnarray}
(\partial^2_{x}-k^2_{y})\phi_{B_{1}j}+E_j\phi_{A_{2}j}&=&E_j^2\phi_{B_{1}j},
\label{eqABCd2a} \\
(\partial^2_{x}-k^2_{y})\phi_{A_{3}j}-E_j\phi_{B_{2}j}&=&-E_j^2\phi_{A_{3}j}.
\label{eqABCd2b}
\end{eqnarray}
\end{subequations}\\
We can  obtain $\phi_{A_{3}j}$ by transforming Eqs. (\ref{eqABCd2a}) and (\ref{eqABCd2b}) into a single sixth-order equation
\begin{widetext}
	\begin{align}\label{eqcubABC}
    \left[ \frac{d^{6}}{dx^{6}}+\left(\alpha_j-3k_y^2\right)\frac{d^{4}}{dx^{4}}-\left(2\alpha_jk_y^2-3k_y^4-\beta_j\right)\frac{d^{2}}{dx^{2}}  -k_y^6
   +\alpha_jk_y^4-\beta_jk_y^2+\omega_j\right]\phi_{A_{3}j}=0,
\end{align}
\end{widetext}
with the  parameters $ \alpha_j=3E_j^2,
\beta_j=E^2_j(3E_j^2-2)$, and
$\omega_j=E_j^2(1-E_j^2)^2$.
Following extensive algebraic manipulations, the solution to Eq. (\ref{eqcubABC}) and therefore the system of equations (\ref{eqdeffABC}) can be expressed as a linear combination of plane waves, represented in terms of matrices specific to each region ($j$=I, III):
\begin{equation}\label{eq9}
\psi_j(x,y)=Q_j M_j(x)C_j e^{ik_{y}y}
\end{equation}
where the three matrices are given by
\begin{align}\label{MatrixQj}
&Q_j=\begin{pmatrix}
   \frac{\lambda_{1j}f_{1j}}{E_j^2g_{1j}} & \frac{\lambda_{1j}g_{1j}}{E_j^2f_{1j}} & \frac{\lambda_{2j}f_{2j}}{E_j^2g_{2j}} & \frac{\lambda_{2j}g_{2j}}{E_j^2f_{2j}} & \frac{\lambda_{3j}f_{3j}}{E_j^2g_{3j}} & \frac{\lambda_{3j}g_{3j}}{E_j^2f_{3j}} \\
  \frac{\lambda_{1j}}{E_jg_{1j}} & -\frac{\lambda_{1j}}{E_jf_{1j}} & \frac{\lambda_{2j}}{E_jg_{2j}} & -\frac{\lambda_{2j}}{E_jf_{2j}} & \frac{\lambda_{3,j}}{E_jg_{3j}} & -\frac{\lambda_{3j}}{E_jf_{3j}}\\
  \frac{\mu_{1j}}{E_jg_{1j}} & -\frac{\mu_{1j}}{E_jf_{1j}} & \frac{\mu_{2j}}{E_jg_{2j}} & -\frac{\mu_{2j}}{E_jf_{2j}} & \frac{\mu_{3j}}{E_jg_{3j}} & -\frac{\mu_{3j}}{E_jf_{3j}} \\
  \frac{\rho_{1j}}{E_j} & \frac{\rho_{j}}{E_j} & \frac{\rho_{2j}}{E_j} & \frac{\rho_{2j}}{E_j} & \frac{\rho_{3j}}{E_j} & \frac{\rho_{3j}}{E_j} \\
  1 & 1 & 1 & 1 & 1 & 1 \\
  \frac{g_{1j}}{E_j} & -\frac{f_{1j}}{E_j} & \frac{g_{2j}}{E_j} & -\frac{f_{2j}}{E_j} & \frac{g_{3j}}{E_j} & -\frac{f_{3j}}{E_j} \\
\end{pmatrix},
\\
&	M_j=\text{diag}(e^{ik_{1j}x}, e^{-ik_{1j}x}, e^{ik_{2j}x}, e^{-ik_{2j}x},  e^{ik_{3j}x}, e^{-ik_{3j}x}),\\
&\label{CjjeqABC}
C_j=(a^{+}_{1j}, b^{-}_{1j}, a^{+}_{2j},  b^{-}_{2j}, a^{+}_{3j}, b^{-}_{3j})^\dagger,
\end{align}
where $g_{nj}=k_{nj}+ik_y$, $f_{nj}=k_{nj}-ik_y$, $\rho_{nj}=E_j^{2}-k_{nj}^{2}-k_y^{2}$, $\mu_{nj}=E_j(\rho_{nj}-1)$, $\lambda_{nj}=\rho_{nj}^2-E^2_j$, and $k_{nj}$ ($n=1, 2, 3$) are the wave vectors along the $x$-direction in regions $j$=I and III, solutions to the following cubic equation
\begin{align}\label{cubeqABC}
\left(k_{nj}^{2}+k^2_y\right)^{3}-\alpha_j\left(k_{nj}^{2}+k^2_y\right)^{2}
+\beta_j\left(k_{nj}^{2}+k^2_y\right)-\omega_j=0,
\end{align}
and the superscript plus/minus indicates the right/left propagation or evanescent states.

\begin{figure}[t!]
	\centering
	\includegraphics[scale=0.4]{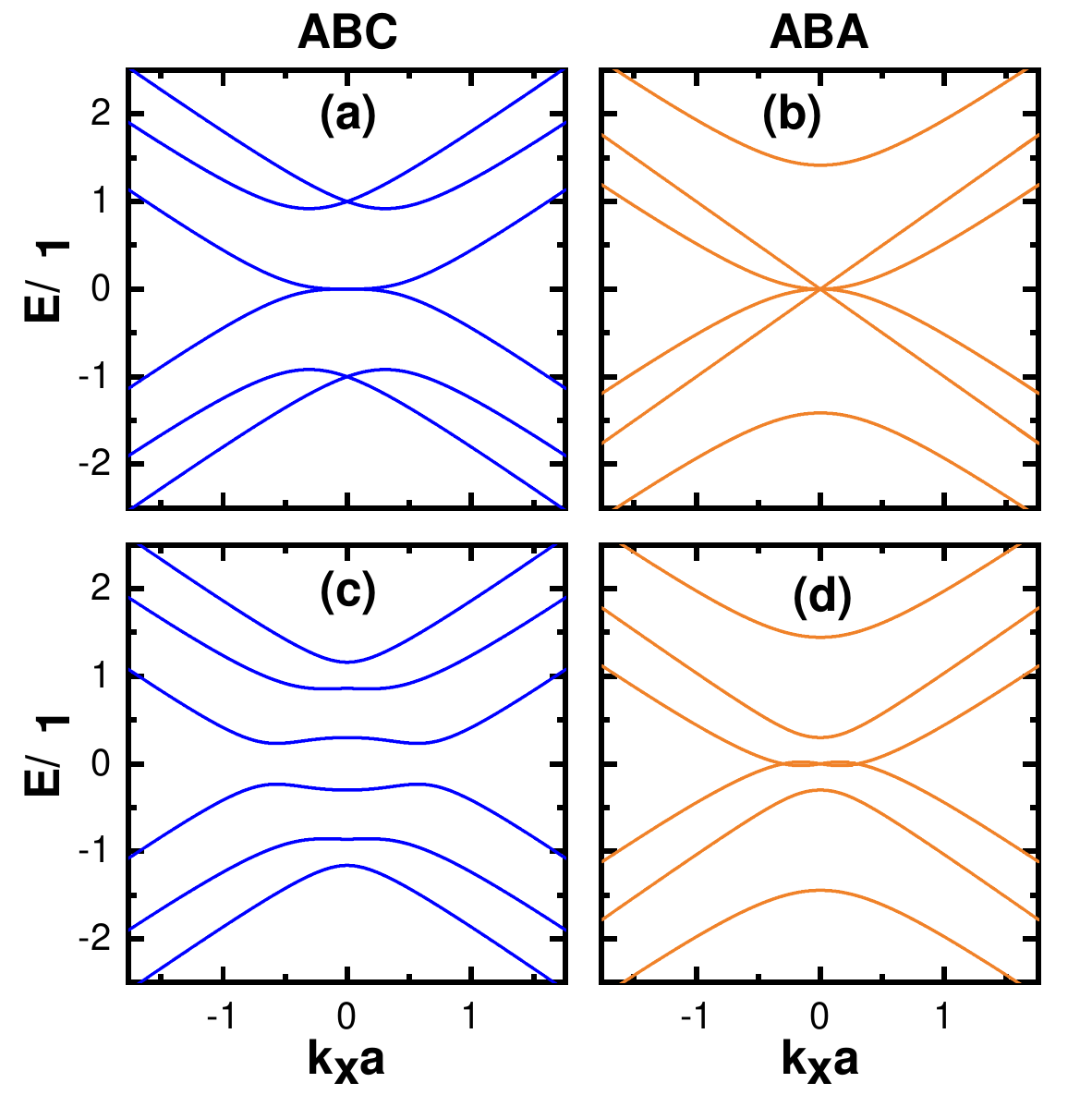} \caption{(Color online) {Energy spectrum as a function of longitudinal wave vector $k_x$ for $k_y=0$ and zero interlayer bias $\delta=0$, with (a): ABC-TLG and (b): ABA-TLG. The chosen non-null value $\delta=0.3\gamma_1$ opens a gap in ABC-TLG (c), but ABA-TLG remains gapless (d).}}
	\label{Energy}
\end{figure}

It is worth noting that in regions $j=\text{I}$ and $\text{III}$ where the potential is zero, the equality $Q_{\text{I}} M_{\text{I}}(x)=Q_{\text{III}} M_{\text{III}}(x)$ holds. As the wave vectors $k_{n,\text{I}}$ are equivalent to those in region III, $k_{n,\text{III}}$, we will simplify our notation in the subsequent analysis by using $k_n$ and omitting the subscript $j$.
It is observed that the energy spectrum of the Hamiltonian (\ref{HamABC}) reveals six bands corresponding to the six wave vectors $\pm k_1$, $\pm k_2$, and $\pm k_3$, solutions of Eq. (\ref{cubeqABC}). This is illustrated in Fig. \ref{Energy}(a) and (c), both without ($\delta=0$) and with bias ($\delta=0.3\gamma_1$). These wave vectors play a crucial role in comprehending the distinct propagation modes across our junction. Specifically, for $E<\gamma_{1}'$ ($\gamma_1'=0.918\gamma_1$) \cite{vanduppen195439,ELMouhafid2200308}, there is only one mode of propagation corresponding to the wave vector $k_1$. However, for $E>\gamma_{1}'$, three propagation modes emerge, associated with the wave vectors $k_1$, $k_2$, and $k_3$, as depicted in Fig. \ref{figbarriermod} (for further details, refer to  \cite{vanduppen195439,ELMouhafid2200308}).

\subsubsection{ABA-TLG in region {\textnormal{II}}}

To obtain the energy spectrum solution for region II (ABA-TLG), one can substitute Eqs. (\ref{HamABA2}) and (\ref{wavefABA}) into  $\mathcal{H}_{ABA}\chi(x)=E\chi(x)$ to derive the coupled equations
\begin{subequations}\label{eqdeffABA}
 \begin{align}
&-i(\partial_{x}+k_{y})\chi_{B_{1}} =\varepsilon\chi_{A_{1}}-\delta\chi_{A_{2}},
 \label{eqsdABA1} \\
&-i(\partial_{x}-k_{y})\chi_{A_{1}} =\varepsilon\chi_{B_{1}}-\delta\chi_{B_{2}}, \label{eqsdeqsdABA2}  \\
&-i(\partial_{x}+k_{y})\chi_{B_{2}}=\varepsilon\chi_{A_{2}}-\delta\chi_{A_{1}}, \label{eqsdeqsdABA3}  \\
&-i(\partial_{x}-k_{y})\chi_{A_{2}}=\varepsilon\chi_{B_{2}}-\delta\chi_{B_{1}}-\sqrt{2}\chi_{A_{3}},
\label{eqsdeqsdABA4} \\
&-i(\partial_{x}+k_{y})\chi_{B_{3}}=\varepsilon\chi_{A_{3}}-\sqrt{2}\chi_{B_{2}},
\label{eqsdeqsdABA5} \\
&-i(\partial_{x}-k_{y})\chi_{A_{3}}=\varepsilon\chi_{B_{3}},
\end{align}
\end{subequations}
where we have set {$\varepsilon=E-V_{0}$}.
The system of coupled first-order differential equations \ref{eqdeffABA}(a-f) can be transformed into a sixth-order differential equation for $\chi_{B_{1}}$ as follows
\begin{widetext}
\begin{align}\label{eqcubABA}
\bigg[\frac{d^{6}}{dx^{6}}-(\omega_1-3k_y^2)\frac{d^{4}}{dx^{4}}-(2\omega_1 k_y^2-3k_y^4-\omega_2)\frac{d^{2}}{dx^{2}} +k_y^6-\omega_1 k_y^4-\omega_2 k_y^2+\omega_3 \bigg]\chi_{B_{1}}=0,
\end{align}
\end{widetext}
with {$\omega_1=3\varepsilon^2+2\delta^2,
\omega_2=3\varepsilon^4-4\varepsilon^2+\delta^4,
\omega_3=\varepsilon^2(\varepsilon^2-\delta^2)(4-\varepsilon^2+\delta^2)$.}
The solution of Eq. (\ref{eqcubABA}) can be written as a linear combination of plane waves
\begin{equation}
\chi_{B_{1}}=\sum^{3}_{n=1}\left(a_{n}e^{iq_n x}+b_{n}e^{-iq_n x}\right),
\end{equation}
where $a_{n}$ and $b_n$ $(n = 1, 2, 3)$ are coefficients of normalization and the wave vectors $q_n$ along the x-direction in each region  are the solutions of the cubic equation:
\begin{widetext}
\begin{align}\label{cubeq}
q_n^{6}-(\omega_1-3k_y^2)q_n^{4}-(2\omega_1 k_y^2-3k_y^4-\omega_2)q_n^{2}+k_y^6-\omega_1 k_y^4-\omega_2 k_y^2+\omega_3=0,
\end{align}
\end{widetext}

In Figs. \ref{Energy}(b) and (d), we present the energy spectrum of the ABA-TLG Hamiltonian (\ref{HamABA1}) both without ($\delta=0$) and with bias ($\delta=0.3\gamma_1$), respectively. The energy spectrum comprises a combination of a pair of nearly linear bands, reminiscent of monolayer graphene, and two pairs of parabolic bands, resembling those of bilayer graphene.
The influence of this energy spectrum behavior significantly affects the transport properties of graphene, as will be discussed in Sec. \ref{RRDD}.

The rest of the spinor component solutions of the system \ref{eqdeffABA}(a-f) are given in terms of the matrices as
\begin{align}\label{MatrixG}
& Q_{\text{II}}=
\begin{pmatrix}
   \frac{\lambda_1f_1}{\alpha_1\alpha_2g_1} & \frac{\lambda_1g_1}{\alpha_1\alpha_2f_1} & \frac{\lambda_2f_2}{\alpha_1\alpha_2g_2} & \frac{\lambda_2g_2}{\alpha_1\alpha_2f_2} & \frac{\lambda_3f_3}{\alpha_1\alpha_2g_3} & \frac{\lambda_3g_3}{\alpha_1\alpha_2f_3} \\
  \frac{\lambda_1}{\alpha_1g_1} & -\frac{\lambda_1}{\alpha_1f_1} & \frac{\lambda_2}{\alpha_1g_2} & -\frac{\lambda_2}{\alpha_1f_2} & \frac{\lambda_3}{\alpha_1g_3} & -\frac{\lambda_3}{\alpha_1f_3}\\
  \frac{\mu_1}{\alpha_1g_1} & -\frac{\mu_1}{\alpha_1f_1} & \frac{\mu_2}{\alpha_1g_2} & -\frac{\mu_2}{\alpha_1f_2} & \frac{\mu_3}{\alpha_1g_3} & -\frac{\mu_3}{\alpha_1f_3} \\
  \frac{\rho_1}{\alpha_1} & \frac{\rho_1}{\alpha_1} & \frac{\rho_2}{\alpha_1} & \frac{\rho_2}{\alpha_1} & \frac{\rho_3}{\alpha_1} & \frac{\rho_3}{\alpha_1} \\
  1 & 1 & 1 & 1 & 1 & 1 \\
  \frac{g_1}{\alpha_1} & -\frac{f_1}{\alpha_1} & \frac{g_2}{\alpha_1} & -\frac{f_2}{\alpha_1} & \frac{g_3}{\alpha_1} & -\frac{f_3}{\alpha_1} \\
\end{pmatrix},
\\
&M_{\text{II}}= \text{diag}(e^{iq_1 x}, e^{-iq_1 x}, e^{iq_2 x}, e^{-iq_2 x}, e^{iq_3 x}, e^{-iq_3 x}),\\
&	\label{Cjjeq}
C_{\text{II}}=(c^{+}_1, d^{-}_1, c^{+}_2, d^{-}_2, c^{+}_3, d^{-}_3)^\dagger,
\end{align}
where {$g_n=q_n+ik_y$, $f_n=q_n-ik_y$, $\rho_n=\alpha_1^{2}-q_n^{2}-k_y^{2}$, $\eta_n=\varepsilon^{2}-q_n^{2}-k_y^{2}$, $\mu_n=\varepsilon\rho_n-\alpha_1$, $\lambda_n=\rho_n\eta_n-\alpha_1\varepsilon$}, $\alpha_1=\varepsilon-\delta$,  $\alpha_2=\varepsilon+\delta$. As a result, the general solution in region II can be written as 
\begin{equation}\label{eq9}
\psi_{\text{II}}(x,y)=Q_{\text{II}} M_{\text{II}}(x)C_{\text{II}}\ e^{ik_{y}y}.
\end{equation}

We will explore how the aforementioned outcomes are applied to ascertain various physical quantities. Our particular emphasis will be on examining the transmission and reflection  probabilities, as well as the conductance.

\subsection{Transmission and conductance}

As we employ the transfer matrix approach \cite{Barbier235408}, our focus lies on the normalization coefficients, which are the components of $C_j$, on both sides of the ABC-ABA-ABC junction. 
In order to provide clarity, we must establish the definition of our spinor within regions I 
\begin{subequations}\label{}
\begin{align}
&\phi_{A_1\text{I}}=\sum^{3}_{n=1}\left(\delta_{s,n}Q_{1,2n-1}e^{ik_nx}+r^{s}_nQ_{1,2n}e^{-ik_nx}\right),\\
&\phi_{B_1\text{I}}=\sum^{3}_{n=1}\left(\delta_{s,n}Q_{2,2n-1}e^{ik_nx}+r^{s}_nQ_{2,2n}e^{-ik_nx}\right),\\
&\phi_{A_2\text{I}}=\sum^{3}_{n=1}\left(\delta_{s,n}Q_{3,2n-1}e^{ik_nx}+r^{s}_nQ_{3,2n}e^{-ik_nx}\right),\\
&\phi_{B_2\text{I}}=\sum^{3}_{n=1}\left(\delta_{s,n}Q_{4,2n-1}e^{ik_nx}+r^{s}_nQ_{4,2n}e^{-ik_nx}\right),\\
&\phi_{A_3\text{I}}=\sum^{3}_{n=1}\left(\delta_{s,n}Q_{5,2n-1}e^{ik_nx}+r^{s}_nQ_{5,2n}e^{-ik_nx}\right),\\
&\phi_{B_3\text{I}}=\sum^{3}_{n=1}\left(\delta_{s,n}Q_{6,2n-1}e^{ik_nx}+r^{s}_nQ_{6,2n}e^{-ik_nx}\right),
\end{align}
\end{subequations}
and III
\begin{subequations}\label{}
\begin{align}
&\phi_{A_1{\text{III}}}=t^s_1Q_{1,1}e^{ik_1x}+t^{s}_2Q_{1,3}e^{ik_2x}+t^{s}_3Q_{1,5}e^{ik_3x},\\
&\phi_{B_1{\text{III}}}=t^s_1Q_{2,1}e^{ik_1x}+t^{s}_2Q_{2,3}e^{ik_2x}+t^{s}_3Q_{2,5}e^{ik_3x},\\
&\phi_{A_2{\text{III}}}=t^s_1Q_{3,1}e^{ik_1x}+t^{s}_2Q_{3,3}e^{ik_2x}+t^{s}_3Q_{3,5}e^{ik_3x},\\
&\phi_{B_2{\text{III}}}=t^s_1Q_{4,1}e^{ik_1x}+t^{s}_2Q_{4,3}e^{ik_2x}+t^{s}_3Q_{4,5}e^{ik_3x},\\
&\phi_{A_3{\text{III}}}=t^s_1Q_{5,1}e^{ik_1x}+t^{s}_2Q_{5,3}e^{ik_2x}+t^{s}_3Q_{5,5}e^{ik_3x},\\
&\phi_{B_3{\text{III}}}=t^s_1Q_{6,1}e^{ik_1x}+t^{s}_2Q_{6,3}e^{ik_2x}+t^{s}_3Q_{6,5}e^{ik_3x},
\end{align}
\end{subequations}
where $\delta_{s,i}$ $(i=1,2,3)$ is the Kronecker symbol, $s=1, 2 ,3$ indicate the mode of propagation or evanescent waves characterized by three distinct wave vectors: $k_1$, $k_2$, and $k_3$. {$Q_{u,v}$} are the elements of the matrix (\ref{MatrixQj}).

By incorporating the suitable boundary condition within the framework of the transfer matrix approach \cite{Barbier235408}, one can derive the transmission and reflection probabilities. The preservation of spinor continuity at the boundaries $(x=0,d)$ yields the components of the vector $C_j$ in regions I and III
\begin{align}\label{eq17}
&C^{s}_{\text{I}}=(\delta_{s,1}, r^{s}_{1},  \delta_{s,2},  r^{s}_{2}, \delta_{s,3}, r^{s}_{3})^\dagger\\
&C^{s}_{\text{III}}=(t^{s}_{1},
0,
t^{s}_{2},
0,
t^{s}_{3},
0)^\dagger,
%
\end{align}
which can be connected via the transfer matrix $M$
\begin{equation}\label{eqTM}
C^{s}_{\text{I}}=M C^{s}_{\text{III}},
\end{equation}
where  $M$ is solution of the continuity equations  
\begin{align}\label{eq19}
 &   Q_{\text{I}} M_{\text{I}}(0)C_{\text{I}}=Q_{\text{II}} M_{\text{II}}(0)C_{\text{II}},\\
  &  Q_{\text{II}} M_{\text{II}}(d)C_{\text{II}}=Q_{\text{III}} M_{\text{III}}(d)C_{\text{III}}.
\end{align}
 Consequently, getting $M$ we can determine the complex coefficients of transmission $t^{s}_{i}$ $(i=1, 2, 3)$ and reflection $r^{s}_{i}$. Given that the velocities of waves scattered through three different modes are not identical, it is advantageous to employ the current density $\vec{\mathbf{J}}$ to calculate the transmission $T$ and reflection $R$ probabilities. This is achieved by 
\begin{equation}\label{jjjj}
\vec{\mathbf{J}}=v_{F}{\Psi}^{\dagger}\vec{\alpha}\Psi,
\end{equation}
where $\Psi$ represents the spinor solutions in regions (I, III), and $\vec{\alpha}$ is a $6\times6$ diagonal matrix with the diagonal elements being the three Pauli matrices $\sigma_{x}$. Utilizing Eq. \eqref{jjjj}, we deduce the incident $\mathbf{J}_{\text{inc}}$, reflected $\mathbf{J}_{\text{ref}}$, and transmitted $\mathbf{J}_{\text{tra}}$ currents that lead to  \cite{vanduppen195439,ELMouhafid2200308}
\begin{equation}\label{eq21}
T=\frac{|\mathbf{J}_{\text{tra}}|}{|\mathbf{J}_{\text{inc}}|}, \quad
R=\frac{|\mathbf{J}_{\text{ref}}|}{|\mathbf{J}_{\text{inc}}|},
\end{equation}
and therefore we end up with
\begin{equation}\label{eq23}
T^{s}_{i}=\frac{A^x_{i,i}}{A^x_{s,s}}|t^{s}_{i}|^{2},\quad
R^{s}_{i}=\frac{A^x_{i,i}}{A^x_{s,s}}|r^{s}_{i}|^{2},
\end{equation}
where $A^x_{i,i}$ and $A^x_{s,s}$ are the elements of the diagonal matrix $\vec{\mathbf{A}}=Q^{\dagger}\vec{\alpha}Q$ consisting of traceless $2\times2$ blocks, and each one corresponds to a propagating mode $s$. 
These expressions can be elucidated as follows: given the presence of six bands, electrons can undergo scattering between them, necessitating consideration of the change in their velocities. Consequently, we identify nine channels in transmission and reflection, corresponding to three modes of propagation: {$k_1$, $k_2$, and $k_3$} solutions of 
Eq.~(\ref{cubeqABC}).

\begin{figure}[ht]
	\centering
	\includegraphics 
	[width=3.3in]
	{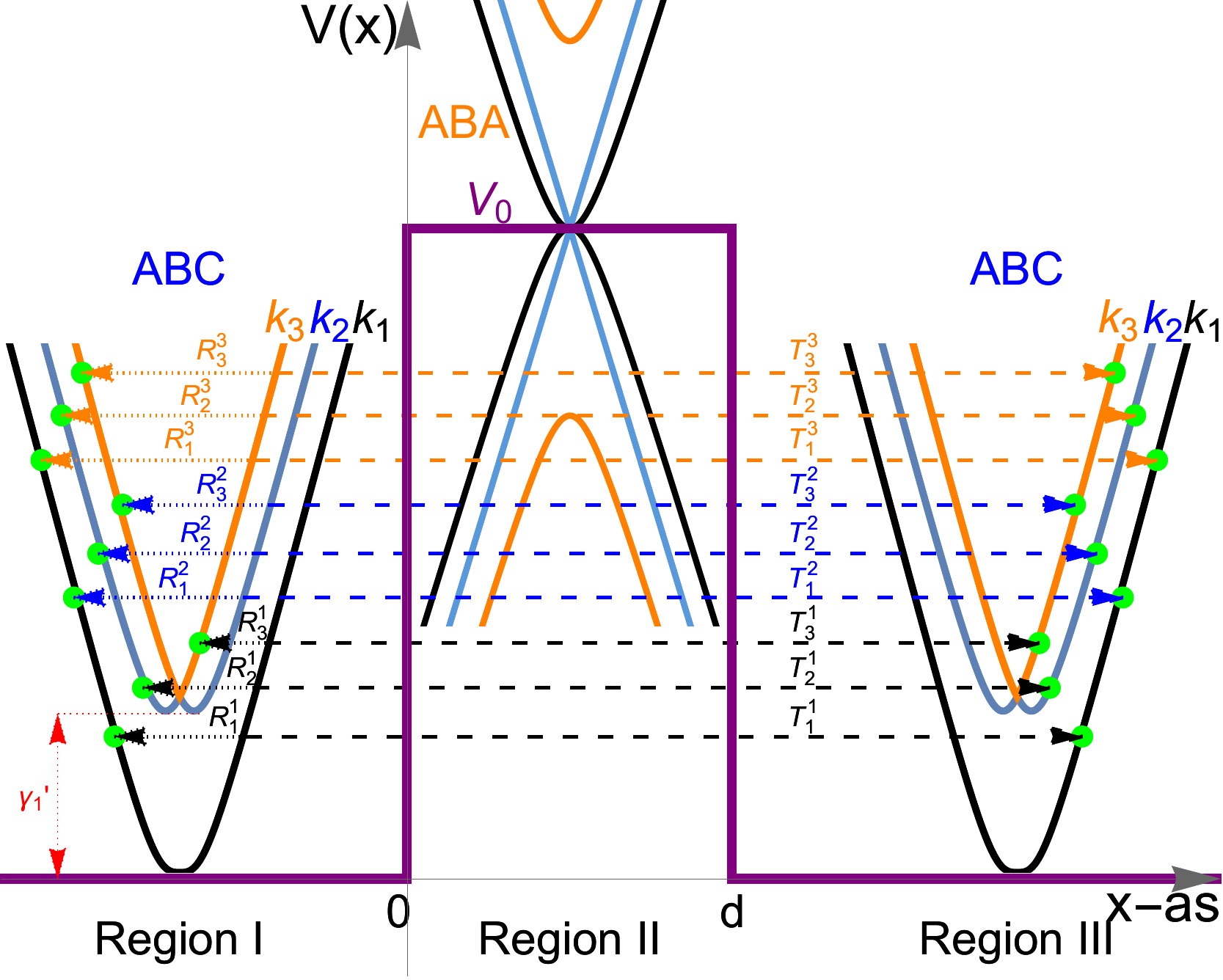}
	\caption{(Color online) Illustration of various modes along with their corresponding transmission and reflection probabilities through the ABC-ABA-ABC junction. Dots and arrows  indicate the energy regions wherein electrons, incident perpendicular to the barrier, will experience transmission or reflection. Here, $\gamma_1'$ represents the minimum energy necessary for the emergence of propagating modes $k_2$ and $k_3$.}\label{figbarriermod}
\end{figure}
In Fig. \ref{figbarriermod}, we illustrate these different modes and the associated transmission and reflection probabilities within the {ABC-ABA-ABC junction}. Specifically, at low energies $(E < \gamma_1')$, a single propagation mode {$k_1$} exists, resulting in one transmission $T^1_1$ and one reflection $R^1_1$ channel across the two conduction bands touching at zero energy on both sides of { the junction}.
Conversely, at higher energies $(E > \gamma_1')$, three propagation modes, {$k_1$, $k_2$, and $k_3$}, give rise to nine transmission $T^{1}_{i}$, $T^{2}_{i}$, $T^{3}_{i}$ $(i=1, 2, 3)$ and reflection $R^{1}_{i}$, $R^{2}_{i}$, $R^{3}_{i}$ channels across the six conduction bands. Concerning transmission, these include three non-scattered channels denoted as $T^{1}_{1}$, $T^{2}_{2}$, and $T^{3}_{3}$ for propagation via $k_1$, $k_2$, and $k_3$, respectively. Additionally, there are six scattered channels where the particle enters through one channel and exits through another. These are denoted as $T^{1}_{3,2}$, $T^{2}_{3,1}$, and $T^{3}_{2,1}$ for scattering from the {$k_1$} band to the {$k_{2,3}$}, from the {$k_2$} band to the {$k_{1,3}$}, and from the {$k_3$} band to the {$k_{1,2}$} bands, respectively. A parallel definition applies to the $R^{1,2,3}_{1,2,3}$ reflection channels. The schematic representation of these nineteen channels is depicted in Fig. \ref{figbarriermod}. It's worth noting that $T^{1,2,3}_{1,2,3}$ and  $R^{1,2,3}_{1,2,3}$ adhere to the following equation:
\begin{equation}
\sum_{{s}=1}^3\left(T^{{s}}_{1,2,3}+R^{{s}}_{1,2,3}\right)=1.
\end{equation}
As illustration, for the lower band in Fig. \ref{figbarriermod}, we have $T^1_{1}+R^1_{1}+T^1_{2}+R^1_{2}+T^1_{3}+R^1_{3}=1.$

Having determined the transmission probabilities, let's examine their impact on the conductance of our system. The conductance can be obtained through the Landauer-B\"{u}ttiker formula \cite{Blanter336} by summing over all channels, resulting in
\begin{equation}\label{eq24}
G(E)=G_{0}\frac{L_y}{2
\pi}\int_{-\infty}^{+\infty}dk_{y}\sum^3_{s,n=1}T^{s}_{n}(E,k_y),
\end{equation}
where $L_y$ represents the width of the sample in the $y$-direction, and $G_0=4e^2/h$, where the factor $4$ accounts for the valley and spin degeneracy in graphene. The subscript $s=1, 2, 3$ denotes the propagation mode. The results obtained will undergo numerical analysis to explore the fundamental characteristics of our system and establish connections with other published findings. Given the nature of our system, the analysis will differentiate between two cases based on band tunneling.

\section{Numerical analysis}\label{RRDD}

\subsubsection{Transmission and reflection probabilities}



\begin{figure}[ht]
	\begin{center}
		\includegraphics[scale=0.38]{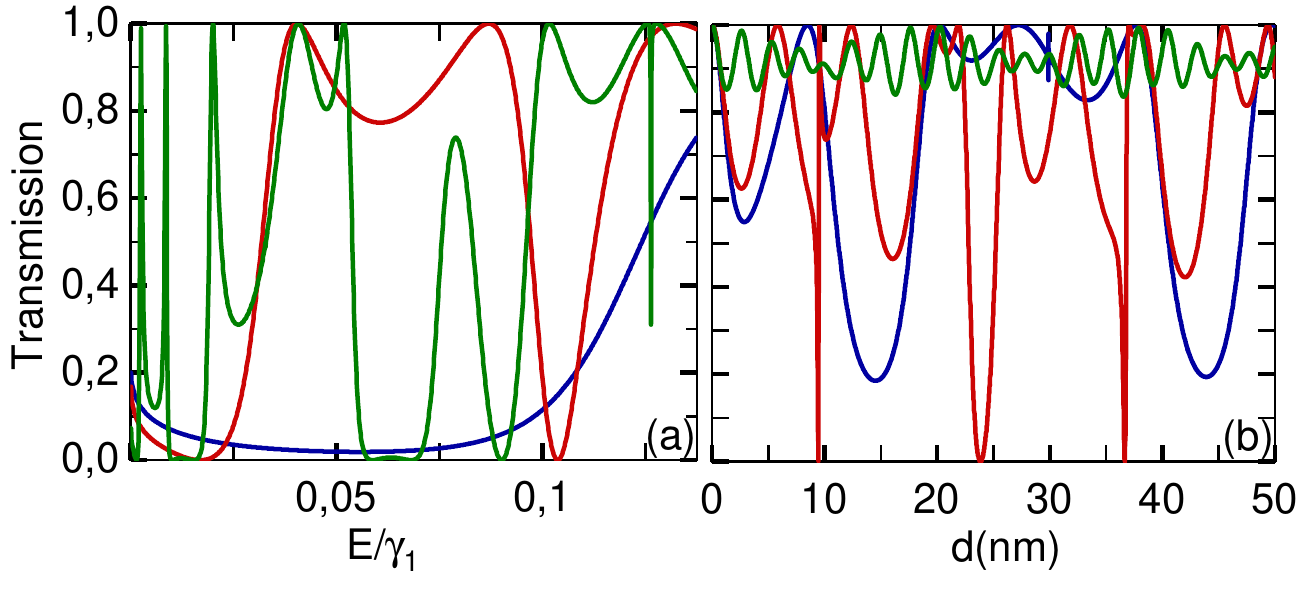}
	\end{center}
	\caption{(Color online) The transmission probability at normal incidence $(k_y=0)$ through an ABC-ABA-ABC junction where $V_{0}=\delta=0$. (a): The energy dependence of the transmission probability for junction width  $d = 10$ nm (blue), $d = 25$ nm (red), and $d = 100 $ nm (green). (b): The width junction dependence of the transmission probability for Fermi energy of  $E =\frac{1}{5} \gamma_1$ (blue), $E =\frac{2}{5}\gamma_1$ (red) and $E =\frac{8}{5}\gamma_1$ (green).} \label{fig0203}  
\end{figure}

To assess transport properties across   the ABC-ABA-ABC junction, we will numerically analyze  our theoretical results. We begin by examining the low-energy case ($E<\gamma_1$), where the corresponding transmission is depicted in Fig. \ref{fig0203} as a function of the incident energy $E$ at normal incidence ($k_y=0)$ for $V_0=\delta=0$. In Fig. \ref{fig0203} (a), the transmission is shown for various junction widths: $d=10$ nm, $d = 25$ nm, and $d=100$ nm. For small $d$ (blue), the transmission is low for $E<0.1\gamma_1$, but it increases and approaches unity for $E>0.1\gamma_1$. As the junction width is increased to $d=25$ nm (red), the transmission is higher even for $E<0.1\gamma_1$, and resonance peaks emerge, surpassing those observed in AB-BLG \cite{Van,BENLAKHOUY2021114835,El Mouhafid2017}. With further increases in junction width, such as $d=100$nm (green), the number of resonance peaks also increases. In Fig. \ref{fig0203} (b), the transmission is plotted against the junction width $d$ for different incident energy values: $E=\frac{1}{5} \gamma_1$ (blue), $E=\frac{2}{5}\gamma_1$ (red), and $E=\frac{8}{5}\gamma_1$ (green). Notably, as the energy increases, the number of resonance peaks also increases, and the transmission exhibits numerous oscillations, surpassing those reported in \cite{Van,BENLAKHOUY2021114835}, even in the absence of $V_0$ and $\delta$.

\begin{figure*}[ht]
	\begin{center}
	\end{center}
	\includegraphics[width=7.1in]{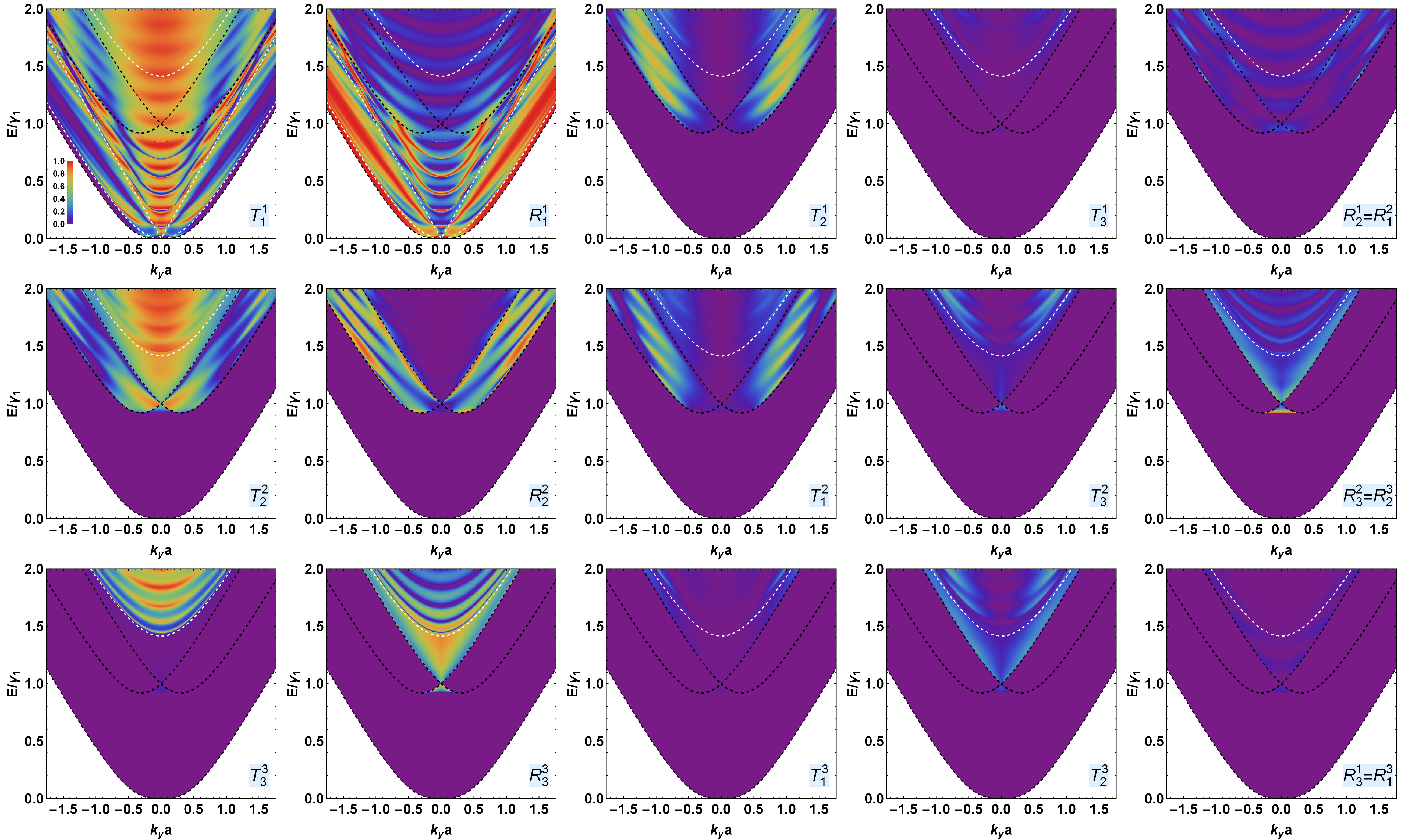}
	\caption{(Color online) Density plot of transmission and reflection probabilities for ABC-ABA-ABC junction with $V_0=\delta=0$ and $d=25\ $nm. The dashed white and black lines represent the band inside (ABA-TLG) and outside (ABC-TLG) the barrier,
		respectively.}
	\label{TransmABC_ABA_ABCV0d0}
	%
\end{figure*}

\begin{figure*}[ht]
	\begin{center}
	\end{center}
	\includegraphics[width=7.1in]{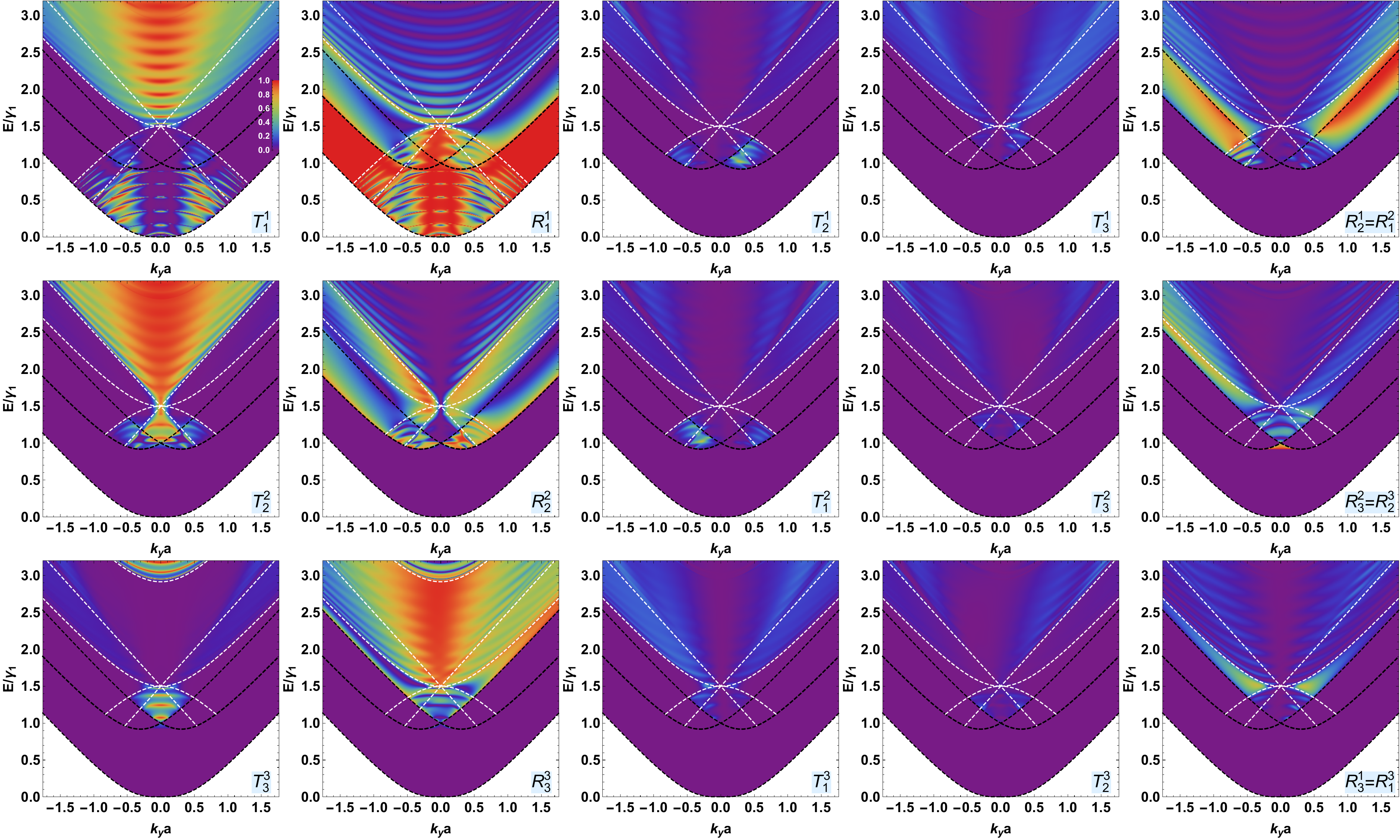}
	\caption{(Color online) The same as in \ref{TransmABC_ABA_ABCV0d0} but now with $V_0=1.5\ \gamma_1$, $\delta=0$.}
	\label{TransmABC_ABA_ABCV15d0}
\end{figure*}

\begin{figure*}[tbh]
	\begin{center}
	\end{center}
	\includegraphics[width=7.1in]{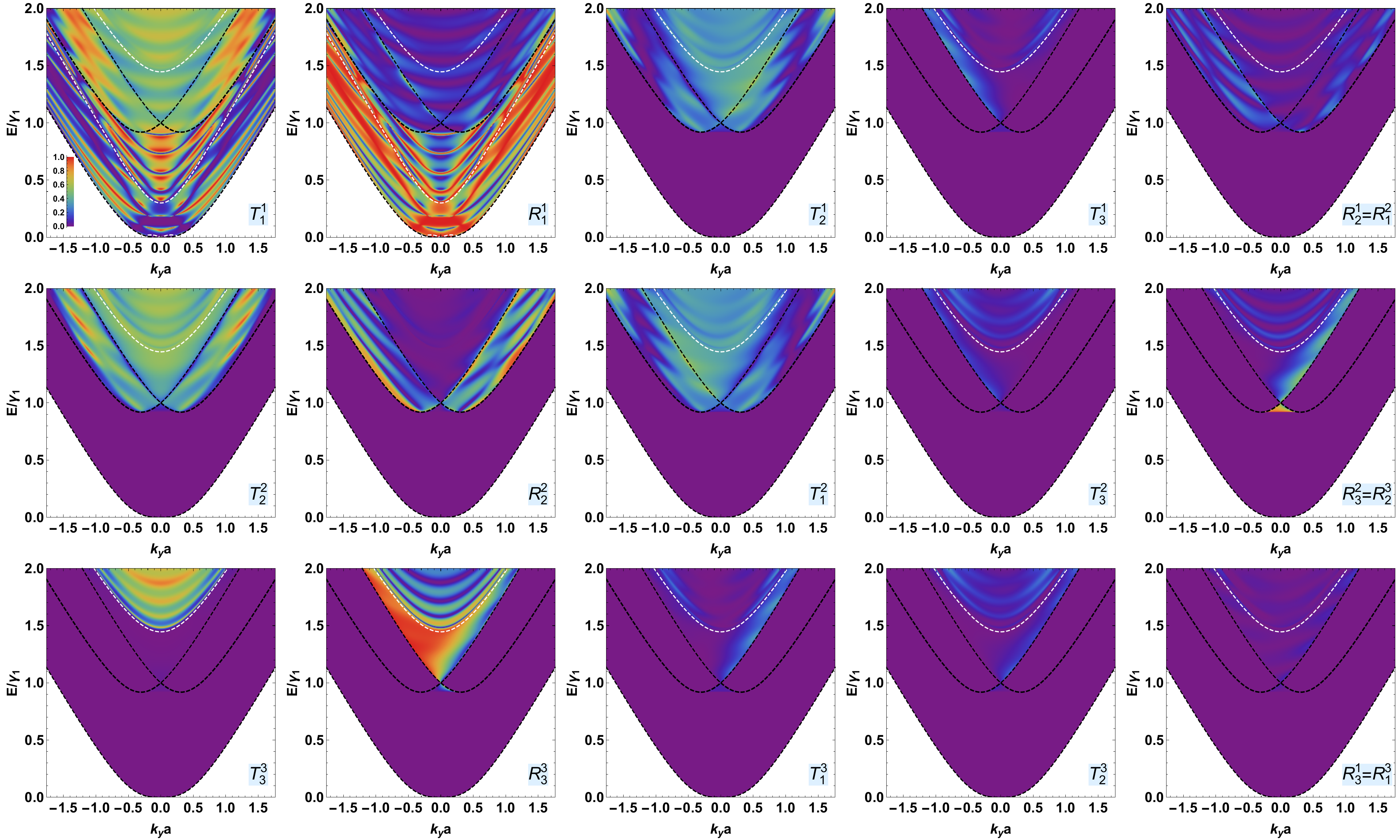}
	\caption{(Color online) The same as in \ref{TransmABC_ABA_ABCV0d0} but now with $V_0=0$, $\delta=0.3\gamma_1$.}
	\label{TransmABC_ABA_ABCV0d03}
\end{figure*}

\begin{figure*}[tbh]
	\begin{center}
	\end{center}
	\includegraphics[width=7.1in]{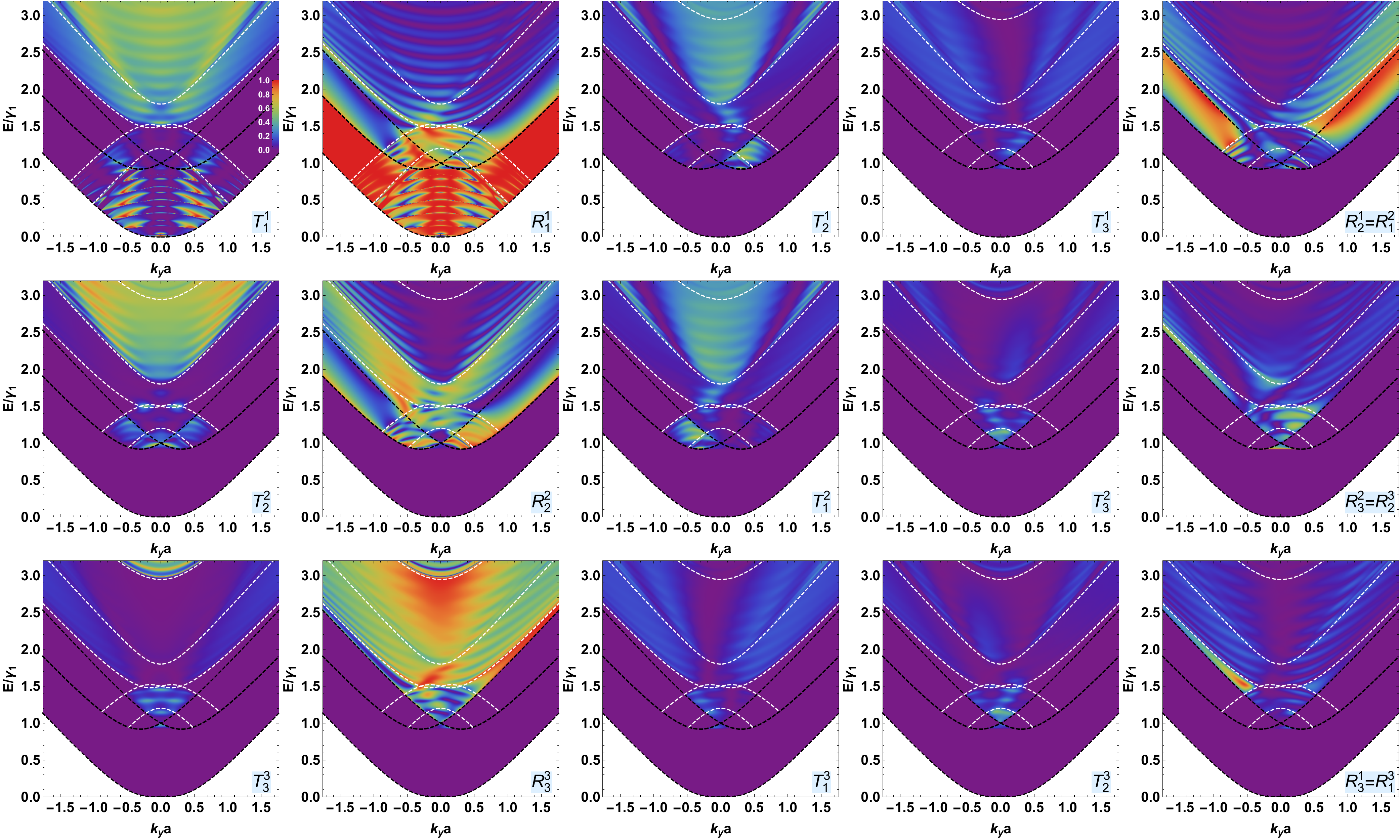}
	\caption{(Color online) The same as in \ref{TransmABC_ABA_ABCV0d0} but now with $V_0=1.5\gamma_1$, $\delta=0.3\gamma_1$ .}
	\label{TransmABC_ABA_ABCV15d03}
\end{figure*}

For energies beyond $\gamma_1'=0.918\gamma_1$ $(E>\gamma_1')$, we depict the transmission and reflection probabilities as a function of energy $E$ and the transverse wave vector $k_y$. The outcomes are presented for $V_0=0$ in Fig. \ref{TransmABC_ABA_ABCV0d0} and for a barrier of height $V_0=1.5\gamma_1$ in Fig. \ref{TransmABC_ABA_ABCV15d0}. It is worth noting that there are nine transmission and nine reflection channels. In the $T^1_1$ channel for $V_0=0$, transmission resonances manifest at normal incidence ($k_y=0$), as shown in Fig. \ref{TransmABC_ABA_ABCV0d0}. This occurrence, recognized as Klein tunneling, is a distinctive feature of the specific junction under consideration and stands in contrast to findings in \cite{Katsnelson}, which require the presence of a potential barrier.
Nevertheless, some narrow anti-resonances are identified for $0<E<\gamma_1'$ at normal incidence, whereas at non-normal incidence, they become broader and even manifest for $E>\gamma_1'$. In the case of the potential barrier $V_0=1.5\gamma_1$ depicted in Fig. \ref{TransmABC_ABA_ABCV15d0}, we observe a reduction in transmission, and anti-Klein tunneling is evidently close to normal incidence in the energy region below $V_0$ in the $T^1_1$ channel. At normal incidence, it occurs for $\gamma_1'<E<V_0$ and at certain energies smaller than $\gamma_1'$. This behavior could be attributed to the presence of BLG-like propagating states inside the barrier, while the minimal transmission at normal incidence may be linked to MLG-like propagating states \cite{Campos,vanduppen195439}. However, transmission resonances emerge at non-normal incidence, where the modes $k_1$ outside and $q_{1,3}$ inside the barrier are coupled, reminiscent of BLG tunneling \cite{Hassane,JELLAL2015149,Redouani,BENLAKHOUY2021114835,Saley}. 
For the $T^2_2$ channel, in both Fig. \ref{TransmABC_ABA_ABCV0d0} and Fig. \ref{TransmABC_ABA_ABCV15d0}, the transmission is null in the energy range $0<E<\gamma_1'$ where the propagating mode $k_2$ is unavailable on the left and right sides of the junction. As $E>\gamma_1'$, the $k_2$ mode becomes accessible, leading to resonances attributed to the propagating modes {$q_{1,2,3}$} inside the junction in Fig. \ref{TransmABC_ABA_ABCV0d0} and $q_{1,3}$ (for $\gamma_1'<E<V_0$) or $q_{1,2}$ (for $E>V_0$) in Fig. \ref{TransmABC_ABA_ABCV15d0}. Notably, the outcome for the $T^2_2$ channel in Fig. \ref{TransmABC_ABA_ABCV15d0} closely resembles the result for the $T^-_-$ channel through the AB-2SL-AB junction with $V_0=1.5 \gamma_1$ \cite{Abdullah}. However, here, we observe non-zero transmission at a specific incidence due to the presence of the propagating mode $q_3$ inside the barrier. In Fig. \ref{TransmABC_ABA_ABCV0d0} for channel $T^3_3$, resonances emerge for $E>\gamma_1'$ where the propagating mode $k_3$ is present outside the junction. Conversely, when $V_0=1.5\gamma_1$, resonances are observed for $\gamma_1'<E<V_0$, in contrast to the result in ABC-TLG \cite{vanduppen195439}, as seen in Fig. \ref{TransmABC_ABA_ABCV15d0}. This difference arises from the presence of the propagating modes $q_1$ and $q_3$ inside the junction, while in \cite{vanduppen195439} the propagating mode $q_3$ is absent. However, no transmission is observed for $0<E<\gamma_1'$ and $V_0<E<2V_0$ due to cloaking \cite{Van}, while a sharp resonance appears for $E>2V_0$.
For $V_0=0$ in Fig. \ref{TransmABC_ABA_ABCV0d0}, the transmission and reflection channels exhibit symmetry with respect to $k_y=0$, akin to the cases of unbiased ABC-TLG \cite{vanduppen195439} and unbiased AB-BLG \cite{Van,Hassane} tunneling. However, for $V_0=1.5\gamma_1$ in Fig. \ref{TransmABC_ABA_ABCV15d0}, the scattered transmission channels, i.e., $T^i_j$ with $i\neq j$, and all reflection channels are no longer symmetric with respect to $k_y=0$, similar to the outcome for the double barrier in biased ABC-TLG \cite{ELMouhafid2200308}. This asymmetrical behavior can be attributed to the non-alignment of ABC-TLG's bands outside and ABA-TLG's bands inside the ABC-ABA-ABC junction due to the presence of a potential barrier \cite{Abdullah}. Additionally, in Fig. \ref{TransmABC_ABA_ABCV15d0}, the transmission decreases in the $T^1_2$ and $T^2_1$ channels when compared with the results in Fig. \ref{TransmABC_ABA_ABCV0d0}, while it remains weak in the other scattered channels in both Fig. \ref{TransmABC_ABA_ABCV0d0} and Fig. \ref{TransmABC_ABA_ABCV15d0}.

With the same parameters as in Fig. \ref{TransmABC_ABA_ABCV0d0}, we present the transmission and reflection probabilities in Fig. \ref{TransmABC_ABA_ABCV0d03} in the presence of an interlayer bias $\delta=0.3 \gamma_1$. It is crucial to acknowledge that within the junction, the interlayer bias transforms ABA-TLG's linear bands into parabolic bands, forming a gap between them while two of the BLG-like bands still touch each other at the Dirac points \cite{LeRoy,vanduppen195439}. Due to the absence of MLG-like propagating states (linear band) inside the junction, the transmission experiences a significant decrease for energy regions greater than $\gamma_1'$ in the $T^i_i$ $(i=1, 2, 3)$, $T^1_2$ and $T^2_1$ channels  compared to those seen in Fig. \ref{TransmABC_ABA_ABCV0d0}. Additionally, the transmission is influenced by $\delta$ in channel $T^1_1$ for energies smaller than $\gamma_1'$. Specifically, for $0<E<\delta$, there is no propagating mode $q_1$ since it has shifted from $E=0$ to $E=\delta$, as observed in \cite{vanduppen195439} for ABA-BLG. Consequently, there are only two sharp resonances resulting from the propagating mode $q_2$. For $\delta<E<\gamma_1'$, the change of propagating mode $q_1$ from MLG-like to parabolic reduces the resonances, as mentioned in \cite{vanduppen195439}. Moreover, in the scattered channels, asymmetry is observed with respect to normal incidence, and here, it arises from the presence of the interlayer bias $\delta$, as shown in \cite{ELMouhafid2200308}. It is important to note that, as in Fig. \ref{TransmABC_ABA_ABCV0d0}, the transmission is weak in channels $T^1_3=T^3_1$ and $T^2_3=T^3_2$. Another noteworthy feature is that in scattered transmission channels, we have $T^i_j(k)=T^j_i(-k)$ as a consequence of the time-reversal symmetry of the system \cite{vanduppen195439}.

In Fig. \ref{TransmABC_ABA_ABCV15d03}, we employ the same parameters as in Fig. \ref{TransmABC_ABA_ABCV15d0}, but with $\delta=0.3\gamma_1$. Similar to Fig. \ref{TransmABC_ABA_ABCV0d03}, we observe a decrease in transmission due to the presence of the interlayer bias, as mentioned previously. Specifically, in the $T^1_1$ and $T^3_3$ channels, no resonance of transmission is found for $E>V_0$, while in the $T^2_2$ channel, only a few sharp resonances occur at non-normal incidence. Additionally, in $T^2_2$ and $T^3_3$ channels, there is no resonance in the region beneath $V_0$, but the transmission differs from zero, contrasting with the result in \cite{ELMouhafid2200308}. In the same energy region, in the $T^1_1$ channel, the result is similar to that obtained in Fig. \ref{TransmABC_ABA_ABCV15d0}. On the other hand, we observe non-zero transmission inside the gap due to the presence of the  propagating mode $q_2$ in both Fig. \ref{TransmABC_ABA_ABCV0d03} and Fig. \ref{TransmABC_ABA_ABCV15d03}. The scattered channels exhibit a similar behavior as in Fig. \ref{TransmABC_ABA_ABCV0d03}, such as asymmetry with respect to $k_y=0$ and $T^i_j(k)=T^j_i(-k)$ as a consequence of the interlayer bias and the time-reversal symmetry of the system \cite{vanduppen195439}. It is worth noting that in ABC tunneling, in the presence of a potential barrier and for a non-zero interlayer bias, a gap is typically observed \cite{ELMouhafid2200308}. However, here, we observe the opposite phenomenon, namely the suppression of the gap when both a potential barrier and an interlayer bias are present.

\subsubsection{Conductance}

\begin{figure*}[ht]
	\centering
	\includegraphics[width=3.5in]{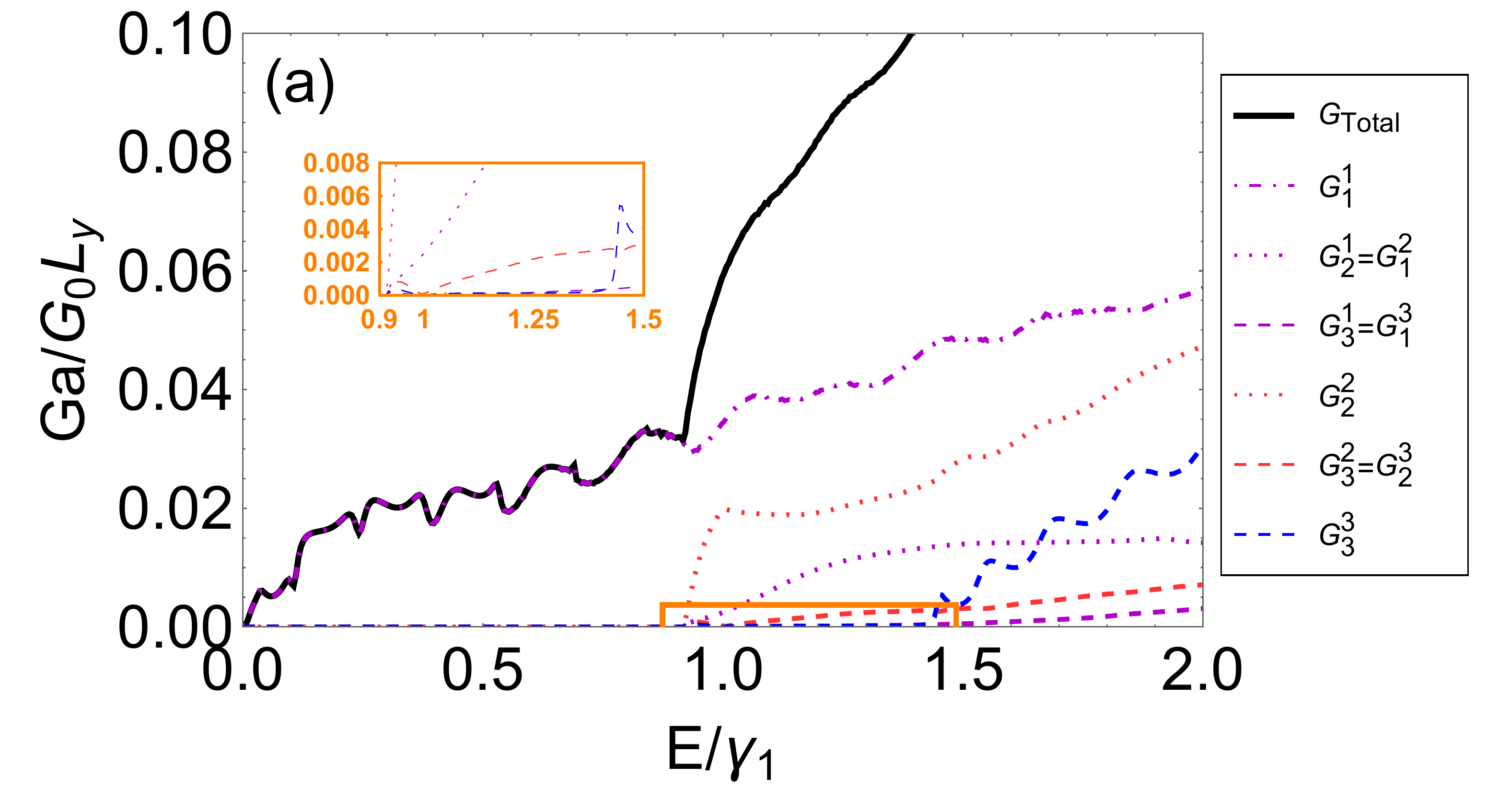}
	\includegraphics[width=3.5in]{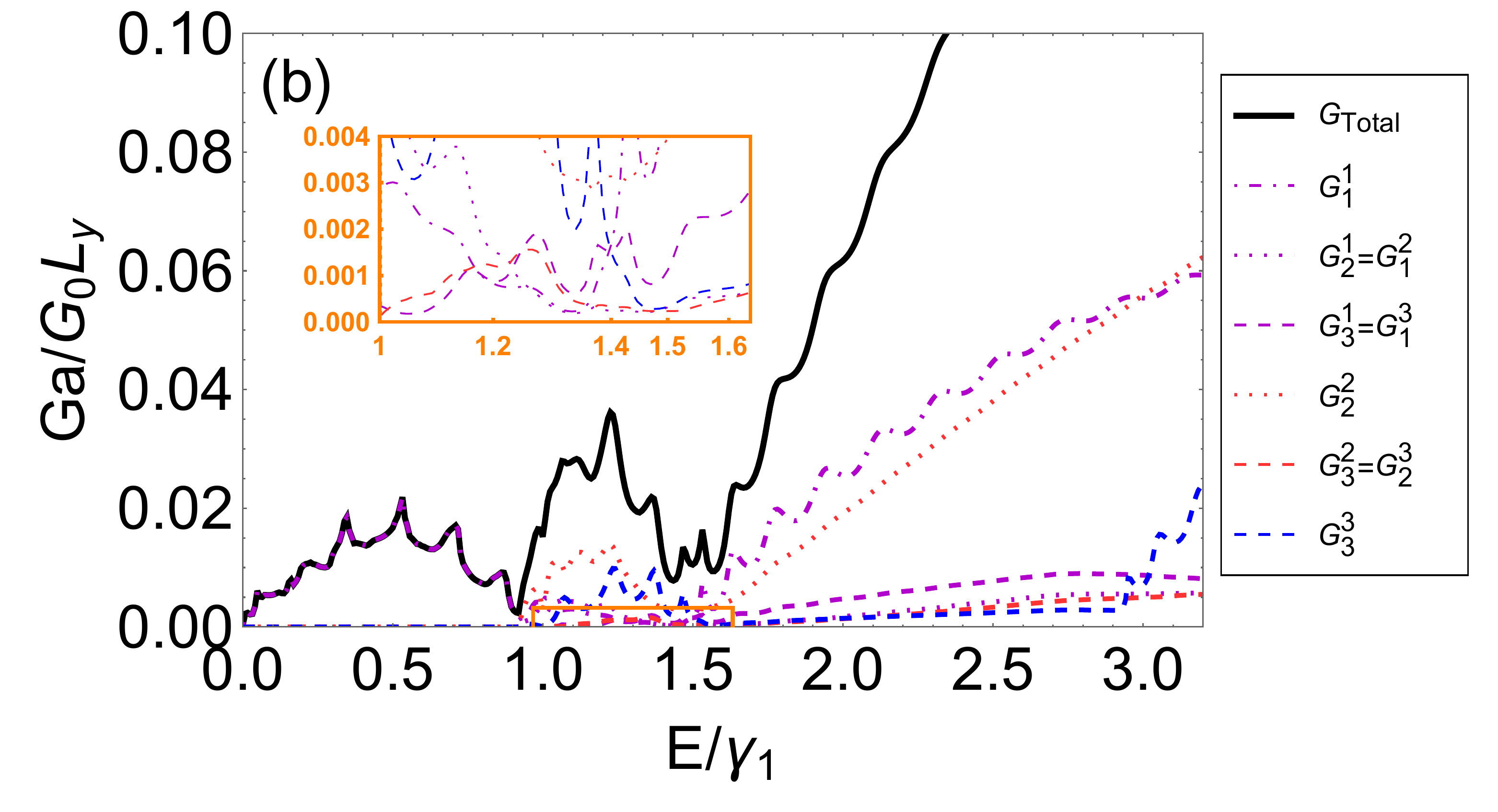}
	\includegraphics[width=3.5in]{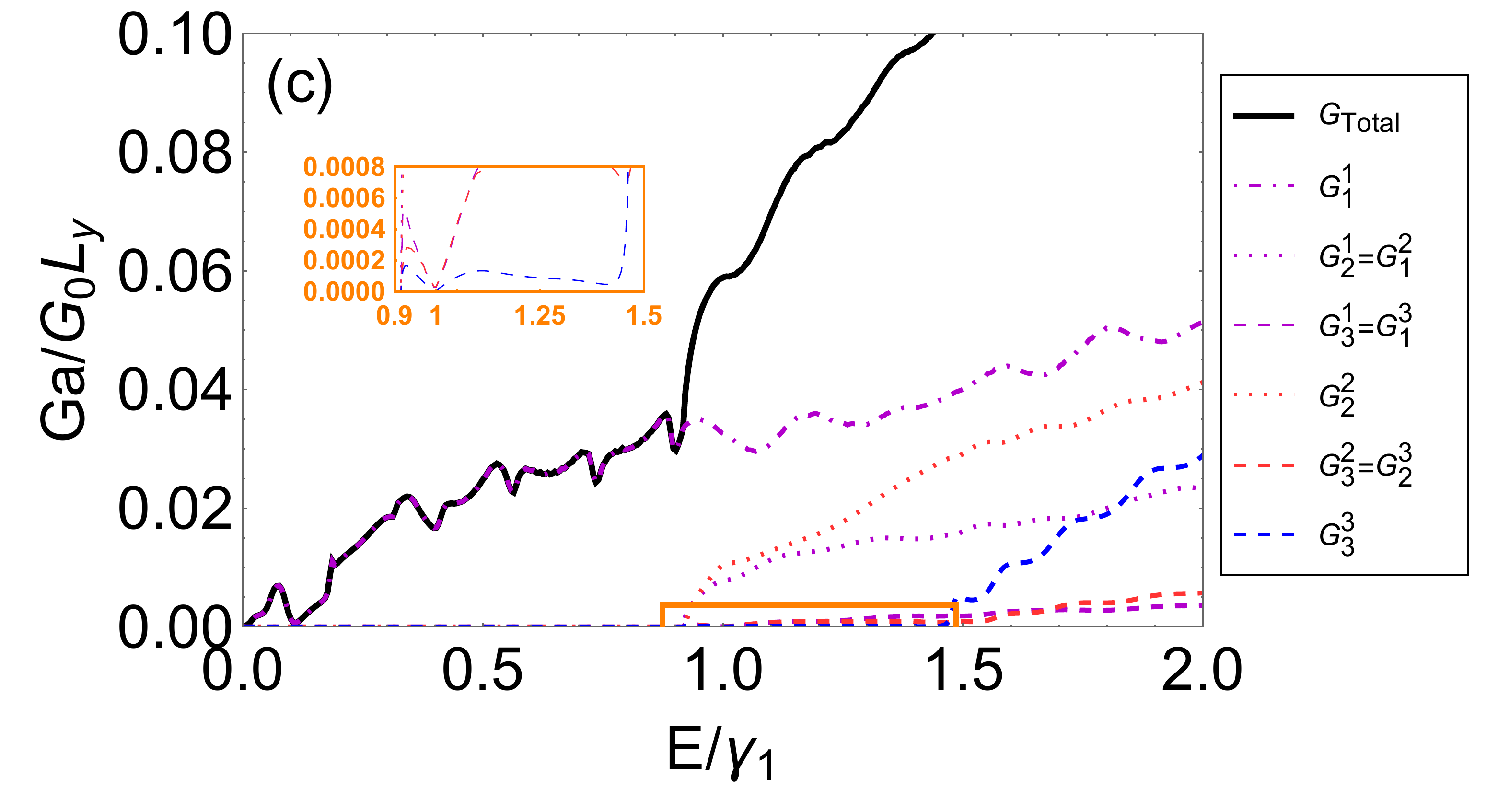}
	\includegraphics[width=3.5in]{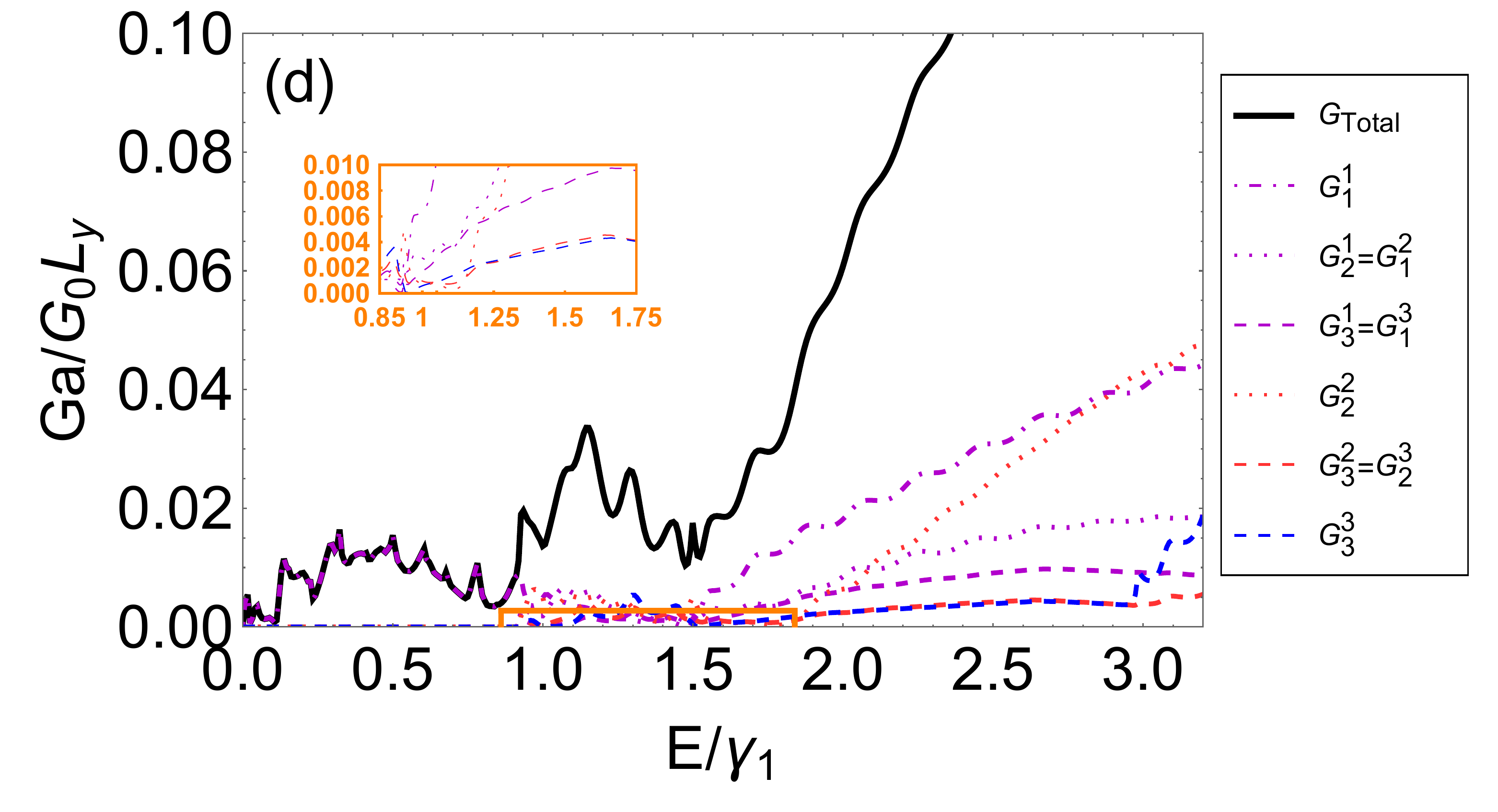}
	\caption{(Color online) Conductance of the  ABC-ABA-ABC junction as a function
		of energy with $d=25\ $nm and for (a) $V_0=\delta=0$, (b) $V_0=1.5\gamma_1$, $\delta=0$, (c) $V_0=0$, $\delta=0.3\gamma_1$ and  (d) $V_0=1.5\ \gamma_1$, $\delta=0.3\gamma_1$.}
	\label{ConductanceABC_ABA_ABC}
\end{figure*}

In Fig. \ref{ConductanceABC_ABA_ABC}, we explore the energy dependence of the conductance through the ABC-ABA-ABC junction under the same conditions as those in the systems studied above. Fig. \ref{ConductanceABC_ABA_ABC}(a) illustrates the conductance associated with the transmission channels in Fig. \ref{TransmABC_ABA_ABCV0d0}. For $E<\gamma_1'$, we have $G_{\text{Total}}=G_1^1$ since in this region electrons can only propagate via mode $k_1$. However, $G^2_2$, $G^1_2=G^2_1$, and $G^2_3=G^3_2$ begin conducting at $E= \gamma_1'$ where more than one propagation mode is available, whereas $G^3_3$ and $G^1_3=G^3_1$ start conducting for $E>\gamma_1'$. The observed equivalence in conductance channels $G^i_j=G^j_i$ ($i \neq j)$ aligns with the results in \cite{vanduppen195439, ELMouhafid2200308, BENLAKHOUY2021114835}. Additionally, a linear-like behavior of $G^1_1$ is noticeable as a consequence of the absence of the potential barrier ($V_0=0$), similar to result observed in \cite{Abdullah}. Furthermore, due to the resonances displayed in Fig. \ref{TransmABC_ABA_ABCV0d0}, the contributions from $G^2_2$, $G^3_3$, and $G^1_2=G^2_1$ significantly increase the total conductance $G_{\text{Total}}$, in contrast to the results seen in \cite{ELMouhafid2200308}. The conductance is notably higher, despite the absence of a potential barrier. However, the contributions from $G^1_3=G^3_1$ and $G^2_3=G^3_2$ are low due to the weak transmission observed in these channels, see  Fig. \ref{TransmABC_ABA_ABCV0d0}. In Fig. \ref{ConductanceABC_ABA_ABC}(b), we plot the conductance associated with the transmission channels in Fig. \ref{TransmABC_ABA_ABCV15d0}. The presence of the potential barrier $V_0=1.5\gamma_1$ alters the conductance behavior in contrast to the result marked in Fig. \ref{ConductanceABC_ABA_ABC}(a). More precisely, it behaves similarly to that observed in \cite{Hassane}. Indeed, the conductance $G_{\text{Total}}$ exhibits minima at $E=\gamma_1'$ and even for $E>\gamma_1'$. Additionally, there is a decrease in $G_{\text{Total}}$, resulting from the reduced transmission found in Fig. \ref{TransmABC_ABA_ABCV15d0}.
In Fig. \ref{ConductanceABC_ABA_ABC}(c), we illustrate the conductance corresponding to Fig. \ref{TransmABC_ABA_ABCV0d03} with $V_0=0$ and $\delta=0.3\gamma_1$. Upon comparing the conductance $G_{\text{Total}}$ with the result in Fig. \ref{ConductanceABC_ABA_ABC}(a), it is evident that, for $E<\gamma_1$, the number of peaks decreases, and $G_{\text{Total}}$ tends to zero at $E=0.1\gamma_1$. This outcome arises from the shift of the propagating mode $q_1$ inside the junction in the $T^1_1$ channel (Fig. \ref{TransmABC_ABA_ABCV0d03}), induced by $\delta$. For $E>\gamma_1'$, there is a decrease in $G^1_1$ and $G^2_2$ compared to the result in Fig. \ref{ConductanceABC_ABA_ABC}(a). This is a consequence of the absence of the MLG-like channel due to the presence of $\delta$, in contrast to Fig. \ref{ConductanceABC_ABA_ABC}(a). In Fig. \ref{ConductanceABC_ABA_ABC}(d), we present the conductance associated with the transmission in Fig. \ref{TransmABC_ABA_ABCV15d03} ($V_0=1.5$ and $\delta=0.3\gamma_1$). We observe that the peaks of $G_{\text{Total}}$ are decreasing in height and increasing in number compared to the result in Fig. \ref{ConductanceABC_ABA_ABC}(b), where $\delta=0$. Additionally, there is a decrease in $G^1_1$, $G^2_2$, and $G^3_3$ channels compared to Fig. \ref{ConductanceABC_ABA_ABC}(b), which is a manifestation of $\delta$. We stress that despite the presence of a potential barrier $V_0$ and an interlayer bias $\delta$, the gap no longer exists in the ABC-ABA-ABC junction, contrary to the ABC stacking studied in \cite{vanduppen195439}. {In fact, this behavior arises because the ABA-TLG remains metallic in the presence of the interlayer bias, as shown in Fig. \ref{Energy}(d). This indicates the existence of propagation states within the junction through which electrons can tunnel, resulting in a non-zero conductance and hence no observable gap.}
Another notable observation is the convergence of $G^2_2$ towards $G^1_1$ with increasing energy in both Figs. \ref{ConductanceABC_ABA_ABC}(b) and (d), which differs from the outcomes reported in \cite{vanduppen195439,ELMouhafid2200308}.

\section{Conclusion}\label{CC}

We have studied the transport properties of Dirac fermions traversing  in the ABC-ABA-ABC  junction. Initially, we determined the eigenvectors and the propagating modes associated with the Hamiltonians $H_{ABC}$ and $H_{ABA}$, describing, respectively, the ABC-TLG and ABA-TLG stackings. Subsequently, employing continuity conditions at the boundaries and the transfer matrix method, we computed the transmission and reflection probabilities. Among the interesting results, we found that even in the absence of a potential barrier (i.e., $V_0=0$), the transport of charge carriers through ABC-ABA-ABC manifests Klein tunneling phenomena at normal incidence. This came as a consequence of the junction under consideration, and it can be understood that there is an induced potential that favors maximum electrons traveling to the transmitted region III.

On the contrary, our findings indicate that the introduction of a potential barrier $V_0$ can effectively diminish Klein tunneling and regulate the transport of charge carriers in our junction. The inclusion of an interlayer bias $\delta$ results in a reduction of resonances for both cases, i.e., $V_0=0$ and $V_0=1.5\gamma_1$. Despite the presence of a potential barrier and an interlayer bias, there is an observed suppression of the gap, in contrast to the outcomes presented in \cite{vanduppen195439}. Additionally, we observed an evident asymmetry in the transmission channels at normal incidence. Moreover, employing Landauer-Buttiker's formula allowed us to compute the conductance of the junction. In the absence of $V_0$ and $\delta$, a notably high conductance was observed. Furthermore, it became evident that the introduction of a potential barrier altered the behavior of the total conductance $G_{\text{Total}}$. It is noteworthy that the suppression of the gap in conductance observed with the application of a barrier potential and an interlayer contrasts with the findings in \cite{vanduppen195439}. {These results suggest that electron transport can be effectively controlled in ABC-ABA-ABC junctions. 
	
	Recent studied have highlighted a fascinating category of materials known as spin-gapless semiconductors. These materials are formed by combining gapless semiconductors with half-metallic ferromagnets \cite{Wang156404, Rani}. In such materials, one spin channel remains gapless while the other spin channel exhibits semiconducting behavior \cite{Wang156404, Rani}. For the ABC-ABA-ABC junction, there is a gap in the ABC region, while the ABA region remains gapless. Therefore, it is plausible to hypothesize that adding magnetic ions to these materials \cite{Wang156404} could potentially lead to the production of spin-gapless semiconductors in such junctions.}


\section*{Author Contributions}

All authors contributed equally to this work. All authors have read and
approved the published version of the manuscript.

\section*{Declaration of competing interest}

The authors declare that they have no known competing financial interests or
personal relationships that might appear to influence the work presented in
this paper.

\section*{Data Availability Statement}
This manuscript has no
associated data or the data will not be deposited. [Authors’
comment: The data that support the findings of this study
are available on request from the corresponding author].

\end{document}